\documentstyle[aps,epsf,eqsecnum,preprint]{revtex}
\begin{document}
\tighten
\draft
\preprint{{\small FTUV/99--52, IFIC/99--54}}
\title{N--quantum approach to quantum field theory
at finite $T$ and $\mu$: the NJL model}
\author{Yasuo Umino and Vicente Vento} 
\address{Instituto de F\'{\i}sica Corpuscular -- C.S.I.C. \\
Departamento de F\'{\i}sica Te\`orica, Universitat de Val\`encia \\
E--46100 Burjassot, Val\`encia, Spain}
\date{\today}
\maketitle
\begin{abstract}
We extend the N--quantum approach to quantum field theory to finite
temperature ($T$) and chemical potential ($\mu$) and apply it to the NJL model.
In this approach the Heisenberg fields are expressed using the Haag expansion
while temperature and chemical potential are introduced simultaneously through a
generalized thermal Bogoliubov transformation. Known mean field results
are recovered using only the first term in the Haag expansion. In addition, we find that
at finite $T$ and in the broken symmetry phase of the model the mean field approximation
can not diagonalize the Hamiltonian. Inclusion of scalar and axial vector diquark channels in the
SU(2)$_{\rm f}$ $\otimes$ SU(3)$_{\rm c}$ version of the model can lead to a lowering 
of the vacuum energy density. We discuss how to go beyond the mean field 
approximation by including higher order terms in the Haag expansion.
\vskip 1cm
\noindent PACS numbers: 03.70.+k, 12.40.$-$y
\vskip 0.25cm
\noindent Keywords: Qunatum Field Theory at Finite $T$ and $\mu$; Haag Expansion;
Thermo Field Theory; NJL model; Color $\bar{3}$ Diquarks 
\end{abstract}

\vfill\eject
\section{Introduction}
\label{intro}
The N--quantum approach (NQA) to quantum field theory developed by
O.W.~Greenberg \cite{gre65,gre94a} is a method to solve the operator equations of 
motion by expanding the 
Heisenberg fields as a complete set of normal--ordered and 
on--shell asymptotic (in-- or out--) fields, including the asymptotic 
fields for bound states. This expansion is known as the Haag expansion 
\cite{haa55}. Although the NQA is yet to be applied to gauge theories successful 
applications thus far indicate that the method can be very powerful in some
cases (see \cite{gre94a} for a summary of the applications and references).
In particular NQA is well suited to describe bound states.

Greenberg and collaborators applied the NQA to study 
dynamical symmetry breaking involving the condensates of composite fields in 
the NJL model \cite{nam61} both without \cite{gre86} 
and with \cite{gre87} isospin. Their method differed from the conventional 
approach to the NJL model \cite{reviews} in several distinct ways:
\begin{itemize}
\item The NQA uses retarded amplitudes (for a Haag expansion using in--fields) with all but one leg 
on--shell instead of off--shell Feynman amplitudes.
\item It allows one to work directly with the fields and states of broken symmetry without 
any reference to those of the unbroken--symmetry vacuum.
\item The explicit structure of the vacuum need not be specified. 
\item Bound state amplitudes are calculated directly rather than 
searching for bound state poles in scattering amplitudes.
\end{itemize}

In both \cite{gre86} and \cite{gre87} it was shown that the well known gap equation for the NJL 
model can be obtained by using only the first term in the Haag expansion in the operator 
equation of motion. Analytical expressions for amplitudes and masses for 
various bound states were found by going one order higher in the expansion and the resulting 
masses agreed with the 
original results \cite{nam61} except for the mass of the vector bound state. 
The form factor of the Nambu--Goldstone boson 
was also examined by extending the Haag expansion to third order in the one--loop 
approximation \cite{gre87}.

Since the NJL model has been used extensively to study the consequences of
chiral symmetry breaking in hot and dense matter \cite{reviews}, it is interesting
to ask whether the NQA to the NJL model can be extended to finite $T$ and $\mu$.
In this letter we shall demonstrate that the answer to this question is
affirmative. We stress that the method presented here
to extend the NQA to finite $T$ and $\mu$ is quite general and is not
restricted to the study of the NJL model. 

Our strategy is based on the extension of the thermal field theory formalism of
Umezawa and collaborators \cite{umezawa} by including a finite chemical potential, and has been
motivated by the application of the NQA to the BCS theory of
superconductivity \cite{gre94b}. The basic idea is to apply a generalized thermal
Bogoliubov transformation to the creation and annihilation operators
of the asymptotic fields in the Haag expansion accompanied by a thermal
doubling of the Hilbert space. In this way temperature and chemical 
potential are introduced simultaneously through the coefficients of the transformation and are
treated as parameters on an equal footing. Quantities of interest
are then obtained through the algebra of thermal creation and
annihilation operators without
specifiying the structure of the thermal vacuum state. 

When the method is applied to the NJL model we recover the known mean field results
for number and scalar densities, total chemical potential, 
gap equation and vacuum energy density using only the first term in the Haag expansion. 
Thus NQA in the lowest order Haag expansion is equivalent to the mean field
approximation usually employed in the studies of the NJL model at finite $T$ and $\mu$. 
Furthermore, in the same approximation, we find that the Hamiltonian can not be diagonalized
at finite $T$ and in the broken symmetry phase of the model due to particle--hole
and anti--particle and anti--hole excitations. This seems to be a general
result valid for all versions of the NJL model at finite $T$ and $\mu$.

In the following section we start by summarizing the pertinent results for the 
U(1) NJL model for $(T, \mu) = (0,0)$ in the NQA. In Section~III the NQA is extended 
to finite $T$ and $\mu$ which is used to recover 
the known mean field results in Section~IV. In the same section we present the off--diagonal 
Hamiltonian and show that it can only vanish in the $T = 0$ or in the
Breit--Wigner phase of the model. We repeat the same excercise in Section~V using the 
SU(2)$_{\rm f}$ $\otimes$ SU(3)$_{\rm c}$ NJL model including the scalar and
axial vector color $\bar{3}$ diquark channels with their respective coupling
constants. By demanding that the total Hamiltonian is diagonalized at $T = 0$ a simple
relation between these coupling constants is found. This relation is then used to show 
how the inclusion of the diquark channels can lower the vacuum energy density of the
many--body system. In the final section we summarize the results and outline the 
procedure for going beyond the mean field approximation.

\section{NQA to the U(1) NJL model: $(T, \mu) = (0,0)$}
\label{sec2}
We first consider the U(1) invariant version of the NJL model 
defined by the Lagrangian
\begin{equation}
{\cal L} = i\bar{\Psi} \partial\!\!\!/ \Psi + 
g_0 \Bigl[ (\bar{\Psi} \Psi )^2 + (\bar{\Psi} i\gamma_5 \Psi)^2  \Bigr].
\label{eq:NJLL}
\end{equation}
The charge conjugation symmetric form of the corresponding field equation 
of motion and Hamiltonian in momentum space are
\begin{eqnarray}
q \!\!\!/ \Psi(q) 
& = & 
-\frac{g_0}{2} \int d^4\!p_1d^4\!p_2d^4\!p_3\; \delta^4(q-p_1-p_2-p_3)  \nonumber \\
&    & 
\otimes \Biggl\{ \biggl[ [\bar{\Psi}(p_1), \Psi(p_2)]_-, \Psi(p_3) \biggr]_+ - 
\biggl[ [\bar{\Psi}(p_1), \gamma_5 \Psi(p_2)]_-, \gamma_5 \Psi(p_3) \biggr]_+ \Biggr\}
\label{eq:HAAGEOM}
\end{eqnarray}
and 
\begin{eqnarray}
{\cal H} 
& = & 
(2\pi)^3 \int d^4\!\bar{p_1} d^4\!\bar{p_2}\; 
\delta^3(\vec{p}_1+\vec{p}_2) e^{-(p_1^0+p_2^0)t} \frac{1}{4} \Bigl[ \bar{\Psi}(p_1), 
\vec{\gamma}\cdot(\vec{p}_1 - \vec{p}_2) \Psi(p_2) \Bigr]_- \nonumber \\
&    & 
- (2\pi)^3 \int d^4\!\bar{p_1} d^4\!\bar{p_2} d^4\!\bar{p_3} d^4\!\bar{p_4}\;
\delta^3(\vec{p}_1+\vec{p}_2+\vec{p}_3+\vec{p}_4) 
e^{-(p_1^0+p_2^0+p_3^0+p_4^0)t} \nonumber \\
&   & 
\;\;\; \otimes \frac{g_0}{4} \Biggl\{ 
\Bigl[\bar{\Psi}(p_1),\Psi(p_2)\Bigr]_- 
\Bigl[\bar{\Psi}(p_3),\Psi(p_4) \Bigr]_- - 
\Bigl[\bar{\Psi}(p_1),\gamma_5 \Psi(p_2) \Bigr]_-
\Bigl[\bar{\Psi}(p_3),\gamma_5 \Psi(p_4) \Bigr]_- \Biggr\},
\label{eq:NJLH}
\end{eqnarray}
where $\bar{\Psi}(p) = \Psi^{\dagger}(-p)\gamma_0$, $[A, B]_\pm = AB \pm BA$ and we shall
adapt the notation $d^n\!\bar{p} \equiv d^n\!p/(2\pi)^3$. When $(T,\mu) = (0,0)$ there are 
no contributions from the exchange (Fock) terms
in the mean field approximation in this model.

For the model defined by Eq.~(\ref{eq:NJLL}) the Haag expansion for the Heisenberg field $\Psi(p)$ 
is given by \cite{gre86},
\begin{eqnarray}
\Psi(q)  
& = & 
\Psi_{\rm IN}(q)\delta_m(q) \nonumber \\
&    & 
+ \sum_k \int d^4\!p_1 d^4\!p_2\; \delta^4(q-p_1-p_2) f^{(k)}(p_1,p_2) 
:\Psi_{\rm IN}(p_1)\delta_m(p_1)F^{(k)}_{\rm IN}(p_2)\delta_{m_k}(p_2): \nonumber \\
&   & 
+ \int d^4\!p_1 d^4\!p_2\; \delta^4(q-p_1-p_2) e(p_1,p_2) C^{-1} 
:\bar{\Psi}_{\rm IN}(p_1)\delta_m(p_1)E_{\rm IN}(p_2)\delta_{m_k}(p_2): \nonumber \\
&   & 
+ \cdot\cdot\cdot.
\label{eq:HAAGEX}
\end{eqnarray}
Here $\Psi_{\rm IN}$ is the asymptotic in--field of $\Psi$ constrained to 
be on the mass--shell by the delta function $\delta_m(p) \equiv \delta^4(p^2 - 
m^2)$. $C = i\gamma_2\gamma_0$ is the charge conjugation operator and the symbol 
$:\hspace{0.5cm}:$ denotes normal ordering. The $k$ summation runs over the $\bar{\Psi} \Psi$, 
$\bar{\Psi} i\gamma_5 \Psi$, and $\bar{\Psi}\gamma_\mu \Psi$ bound states 
with corresponding on--shell in--fields $F^{(k)}_{\rm IN}$ and amplitudes 
$f^{(k)}(p_1,p_2)$. Similarly $e(p_1,p_2)$ is the amplitude for the $\Psi\Psi$ bound state with an 
on--shell asymptotic field $E_{\rm IN}$. The ellipses represent higher order contributions 
involving terms with three or more products of $\Psi_{\rm IN}$, $\bar{\Psi}_{\rm IN}$ 
and asymptotic bound state fields together with corresponding amplitudes. An example of the third 
order contribution is given in \cite{gre87}. 

As shown in \cite{gre86}, inserting the first term in the above expansion into the operator 
equation of motion Eq.~(\ref{eq:HAAGEOM}) yields the well known gap equation
\begin{equation}
m = 4g_0 \int d^3\!\bar{p}\;\frac{m}{\omega_p}
\label{eq:GAP0}
\end{equation}
with $\omega_p = \sqrt{{|\vec{p}\,|^2 + m^2}}$. Note that the asymptotic 
mass $m$ plays the role of the chiral gap. There are three solutions to this 
gap equation; $m=0$ corresponding to the phase with unbroken chiral symmetry 
and $m =\pm m_0$ for the broken symmetry phase. The  
vacuum energy densities for the two phases are
\begin{eqnarray}
\frac{1}{V}\langle {\cal H} \rangle^{m=0} 
& =  &
-2 g_0\int d^3\!\bar{p}d^3\!\bar{q} 
-2 \int d^3\!\bar{p}\; |\vec{p}\,| \label{eq:EVAC0} \\
\frac{1}{V}\langle {\cal H}  \rangle^{m=\pm m_0} 
& =  & 
-2 g_0\int d^3\!\bar{p}d^3\!\bar{q} 
- \int d^3\!\bar{p}\; \frac{2|\vec{p}\,|^2 + m^2}{\omega_p} 
\label{eq:EVACM}
\end{eqnarray}

To facilitate comparison with known results for finite $T$ and $\mu$ 
we shall use the non--covariant regularization scheme and cut off 
the three momentum integral at $|\vec{p}\,| = \Lambda$. Then, as is
well known, dynamical symmetry 
breaking occcurs only for those values of the coupling constant satisfying
the relation $\pi^2/\Lambda^2 < g_0$. For these values of $g_0$ the vacuum 
energy density of the broken symmetry phase is lower than that of the 
unbroken phase, indicating that the phase with chiral condensates is the 
energetically favored phase.

\section{Extension to finite $T$ and $\mu$}
\label{sec3}
In this work we shall explore the consequences of using only the first term 
in Eq.~(\ref{eq:HAAGEX}) to describe the NJL model at finite $T$ and $\mu$. At first sight 
this seems to be a trivial approximation, but as we shall see it turns out not to be the 
case. Extension of the present work to include the second order terms will 
be discussed below.

The lowest order term in the Haag expansion 
for $\Psi$ is \footnote{We use the conventions and normalizations 
of Itzykson and Zuber \cite{IZ}.}
\begin{equation}
\Psi_{\rm IN}(q) \delta_m(q) = \frac{1}{(2\pi)^3}\frac{m}{\omega_q} \sum_{s=\pm} 
\biggl[ b(\vec{q},s) u(\vec{q},s) \delta(q^0-\omega_q) + d^{\dagger}(-\vec{q},s) v(-\vec{q},s) 
\delta(q^0+\omega_q) \biggr].
\label{eq:NMODES}
\end{equation}
Because this term alone reproduces the gap equation Eq.~(\ref{eq:GAP0}) 
and diagonalizes the Hamiltonian, the operators $b$ and $d$ correspond to quasi--particle and
quasi--anti--particle annihilation operators, 
respectively, which annihilate the interacting vacuum state at zero temperature 
and chemical potential $|{\cal G}(0,0)\rangle$, 
\begin{equation}
b(\vec{q},s) |{\cal G}(0,0)\rangle = d(\vec{q},s) |{\cal G}(0,0)\rangle = 0.
\label{eq:AVAC0}
\end{equation}
Thus using the first order term in the Haag expansion is equivalent to 
subjecting the creation and annihilation operators in the Heisenberg 
field $\Psi$ to a Bogoliubov transformation into quasi--particle and 
quasi--anti--particle basis.

However, the $b$ and $d$ operators do not annihilate the interacting vacuum
state at finite $T$ and $\mu$ denoted as $|{\cal G}(T,\mu)\rangle$. In order to 
construct operators that annihilate $|{\cal G}(T,\mu)\rangle$ we apply a generalized 
thermal Bogoliubov transformation to the $b$ and $d$ operators as well as to the 
annihilation operators corresponding to quasi--holes $\tilde{b}$ and 
quasi--anti--holes $\tilde{d}$ as follows
\begin{eqnarray}
b(\vec{q},s) 
& = & 
\alpha_q B(\vec{q},s) - s \beta_q \tilde{B}^{\dagger}(-\vec{q},s)
\label{eq:BT1} \\
\tilde{b}(\vec{q},s) 
& = & 
\alpha_q \tilde{B}(\vec{q},s) + s \beta_q B^{\dagger}(-\vec{q},s)
\label{eq:BT2} \\
d(\vec{q},s) 
& = & 
\gamma_q D(\vec{q},s) - s \delta_q \tilde{D}^{\dagger}(-\vec{q},s)
\label{eq:BT3} \\
\tilde{d}(\vec{q},s) 
& = & 
\gamma_q \tilde{D}(\vec{q},s) + s \delta_q D^{\dagger}(-\vec{q},s)
\label{eq:BT4}
\end{eqnarray}
Here $B$ and $\tilde{B}$ annihilate a quasi--particle and a quasi--hole
at finite $T$ and $\mu$, respectively, and $D$ and 
$\tilde{D}$ are the thermal annihilation operators for quasi--anti--particles 
and quasi--anti--holes, respectively. These operators annihilate 
the interacting thermal vacuum state {\em for each $T$ and $\mu$}. 
\begin{equation}
B(\vec{q},s) |{\cal G}(T,\mu)\rangle 
=\tilde{B}(\vec{q},s) |{\cal G}(T,\mu)\rangle 
= D(\vec{q},s) |{\cal G}(T,\mu)\rangle 
=\tilde{D}(\vec{q},s) |{\cal G}(T,\mu)\rangle 
= 0
\label{eq:AVAC}
\end{equation}
In the NQA it is not necessary to specify the structure of
$|{\cal G}(T,\mu)\rangle$. We only have to assume that it is annihilated by the 
annihilation operators in the Haag 
expansion. The thermal doubling of the Hilbert space accompanying the thermal
Bogoliubov transformation is implicit in Eq.~(\ref{eq:AVAC})
where a ground state which is annihlated by thermal operators 
$B$, $\tilde{B}$, $D$ and $\tilde{D}$ is defined.

In addition, thermal operators satisfy the Fermion anti--commutation relations
\begin{eqnarray}
(2\pi)^3 \frac{\omega_p}{m} \delta^3\left( \vec{p} - \vec{q}\right) 
\delta_{s_1s_2} 
& = &
\biggl[B^{\dagger}(\vec{p},s_1), B(\vec{q},s_2) \biggr]_+ =
\biggl[D^{\dagger}(\vec{p},s_1), D(\vec{q},s_2) \biggr]_+ \nonumber \\
& = &
\biggl[\tilde{B}^{\dagger}(\vec{p},s_1), \tilde{B}(\vec{q},s_2) \biggr]_+ =
\biggl[\tilde{D}^{\dagger}(\vec{p},s_1), \tilde{D}(\vec{q},s_2) \biggr]_+
\label{eq:NORM}
\end{eqnarray}
with vanishing anti--commutators for the remaining combinations.
The coefficients of the transformation are $\alpha_q = \sqrt{1-n_q^-}$,
$\beta_q = \sqrt{n_q^-}$, $\gamma_q = \sqrt{1-n_q^+}$ and
$\delta_q = \sqrt{n_q^+}$, where 
$n_q^{\pm}= [e^{(\omega_q \pm \mu)/(k_B T)}+1]^{-1}$ are the Fermi 
distribution functions for particles and anti--particles.
They are chosen so that the total particle number densities are given by
\begin{eqnarray}
n_q^- 
& = & \frac{1}{V} \frac{m}{\omega_q}
\langle {\cal G}(T,\mu)|b^{\dagger}(\vec{q},s)b(\vec{q},s)|{\cal G}(T,\mu)\rangle  \\
n_q^+& = & \frac{1}{V} \frac{m}{\omega_q}
\langle {\cal G}(T,\mu)|d^{\dagger}(\vec{q},s)d(\vec{q},s)|{\cal G}(T,\mu)\rangle
\label{eq:DIST}
\end{eqnarray}
Hence in this approach temperature and chemical potential are introduced through
the coefficients of the thermal Bogoliubov transformation and are treated as parameters.

The first order Haag expansion for $\Psi$ extended to finite $T$ and $\mu$ is
\begin{eqnarray}
\Psi_{\rm IN}(q) \delta_m(q) 
& = &
 \frac{1}{(2\pi)^3}\frac{m}{\omega_q} \sum_{s=\pm} \Biggl\{
\biggl[ \alpha_q B(\vec{q},s) - s \beta_p \tilde{B}^{\dagger}(-\vec{q},s)\biggr] 
u(\vec{q},s) \delta(q^0-\omega_q)  \nonumber \\
&    &
\;\;\;\;\;\;\;\;\;\;\;\;\;\;\;\;\;\;\;\; +
\biggl[ \gamma_q D^{\dagger}(-\vec{q},s) - s \delta_q 
\tilde{D}(\vec{q},s) \biggr] 
v(-\vec{q},s) \delta(q^0+\omega_q) \Biggr\}.
\label{eq:HAAGTMU}
\end{eqnarray}
Note that this extension is independent of the structure of the Lagrangian and 
is therefore not restricted to the NJL model. If bosonic fields are
present, as in the second order terms in Eq.~(\ref{eq:HAAGEX}), a thermal
Bogoliubov transformation for bosonic 
creation and annihilation operators must be introduced together
with a redefinition of the thermal ground state so that it is
annihilated by both fermionic and bosonic annihilation operators. 
In the following sections 
we shall use the above ansatz for the $\Psi$ field to study the NJL model at 
finite $T$ and $\mu$. 

\section{NQA to the U(1) NJL model: $(T, \mu) \neq (0,0)$}
\label{sec4}

Before extending the U(1) NJL model to finite $T$ and $\mu$ we first
calculate the number (${\cal N}$) and scalar (${\cal S}$) densities using the
ansatz given in Eq.~(\ref{eq:HAAGTMU}). It is a 
straightforward exercise to show that the number density is given by
\begin{eqnarray}
{\cal N} 
& = & 
\frac{1}{V}(2\pi)^3 \int d^3\!\bar{p_1}d^3\!\bar{p_2}\; 
\delta^3(\vec{p}_1 + \vec{p}_2)\langle {\cal G}(T,\mu)
|\frac{1}{2}\Bigl[\Psi^{\dagger}(p_1),\Psi(p_2)\Bigr]_-|
{\cal G}(T,\mu)\rangle \nonumber \\
& = &
2\int d^3\!\bar{p}\;\Bigl(\beta_p^2 - \delta_p^2 \Bigr)
\end{eqnarray}
while the scalar density is found to be
\begin{eqnarray}
{\cal S} 
& = & 
\frac{1}{V}
(2\pi)^3 \int d^3\!\bar{p_1}d^3\!\bar{p_2}\; 
\delta^3(\vec{p}_1 + \vec{p}_2)\langle {\cal G}(T,\mu)
|\frac{1}{2}\Bigl[\bar{\Psi}(p_1),\Psi(p_2)\Bigr]_-|
{\cal G}(T,\mu)\rangle \nonumber \\
& = &
-2\int d^3\!\bar{p}\;\frac{m}{\omega_p}\Bigl(1 - \beta_p^2 - \delta_p^2 \Bigr)
\end{eqnarray}

We now insert a bare chemical potential term $\mu_0\bar{\Psi}\gamma_0\Psi$ 
in the Lagrangian Eq.~(\ref{eq:NJLL}) and use Eq.~(\ref{eq:HAAGTMU}) in the 
corresponding equation of motion and renormal order. Keeping only the 
linear terms in the thermal field operators, one finds that
the equation of motion is given by
\begin{equation}
\Bigl(m + \mu_0\gamma_0 \Bigr) \Psi(q) =
\Biggl\{\Bigl[ 4 g_0 \int d^3\!\bar{p}\;\frac{1}{\omega_p} 
\Bigl(1 - \beta_p^2 - \delta_p^2 \Bigr)\Bigr]m 
+ \Bigl(g_0 {\cal N} \Bigr) \gamma_0 \Biggr\}\Psi(q)
\label{eq:EOMTMU}
\end{equation}
Whereas the exchange terms did not contribute at $(T,\mu)=(0,0)$, when introducing 
$T$ and $\mu$ through the thermal Bogoliubov transformation
one finds contributions to the equation of motion from both the direct and exchange
terms. In fact, the exchange terms generate a contribution to the chemical
potential proportional to the number density. From Eq.~(\ref{eq:EOMTMU}) we see that
the total chemical potential is given by $\mu = \mu_0 - g_0 {\cal N}$.

The gap equation for $(T,\mu)\neq(0,0)$ is found by equating the scalar 
terms in Eq.~(\ref{eq:EOMTMU}). The results is
\begin{equation}
m = 4 g_0 \int d^3\!\bar{p}\;\frac{m}{\omega_p} 
\Bigl(1 - \beta_p^2 - \delta_p^2 \Bigr)
\label{eq:GAPTMU}
\end{equation}
and reduces to Eq.~(\ref{eq:GAP0}) in the $(T,\mu)\rightarrow (0,0)$ limit. To
obtain the vacuum energy density $\epsilon(T,\mu)$ we use Eq.~(\ref{eq:NJLH}) (plus the
$\mu_0$ term) and keep terms with two contractions after renormal ordering. 
We find
\begin{eqnarray}
\epsilon(T,\mu)
& = & 
\frac{1}{V} \langle {\cal G}(T,\mu)|H|{\cal G}(T,\mu)\rangle \nonumber \\
& = & 
-\Biggl\{2g_0\int d^3\!\bar{p}d^3\!\bar{q} + \frac{1}{2}g_0 {\cal N}^2
+ 2\int d^3\!\bar{p}\;\omega_p\; (1 - \beta_p^2 - \delta_p^2) 
-\frac{m^2}{4g_0} + \mu_0{\cal N} \Biggr\} 
\label{eq:INTE} 
\end{eqnarray}

Above expressions for the number and scalar densities, total
chemical potential, the gap equation and the
vacuum energy density all agree with the standard mean field results for the U(1) 
NJL model \cite{reviews}. We note that unlike in the variational method 
the explicit structure of the thermal vacuum state need not be specified when obtaining these 
results. Instead they were determined by the algebra of the thermal fields
defined by Eq.~(\ref{eq:AVAC}) and (\ref{eq:NORM}). We now turn to the discussion 
of the diagonalization of the model Hamiltonian.

In order to obtain the off--diagonal terms in the Hamiltonian, denoted as 
${\cal H}_{{\rm OD}}$, to second order in the thermal creation operators 
we keep terms involving one contraction when renormal ordering 
Eq.~(\ref{eq:NJLH}) (again, with the $\mu_0$ term), and exploit the fact that the thermal 
annihilation operators annihilate the thermal vacuum state. 
Using the gap equation Eq.~(\ref{eq:GAPTMU}) the result can be written as 
\begin{eqnarray}
{\cal H}_{{\rm OD}} 
& = & 
m \sum_s s \int d^3\!\bar{p}\;\Biggl\{ 
\biggl[\alpha_p\beta_p \tilde{B}^{\dagger}(\vec{p},s) B^{\dagger}(-\vec{p},s) 
+ \gamma_p\delta_p \tilde{D}^{\dagger}(-\vec{p},s) D^{\dagger}(\vec{p},s) \biggr] \nonumber \\
&   & 
\;\;\;\;\;\;\;\;\;\;\;\;\;\;\; - 
\frac{\mu}{\omega_p} 
 \biggl[\alpha_p\beta_p \tilde{B}^{\dagger}(\vec{p},s) B^{\dagger}(-\vec{p},s) 
 - \gamma_p\delta_p \tilde{D}^{\dagger}(-\vec{p},s) 
D^{\dagger}(\vec{p},s) \biggr]\Biggr\} 
\label{eq:HOFF1}
\end{eqnarray}
We see that the off--diagonal terms consist of particle--hole
and anti--particle--anti--hole excitations coupled to the spin triplet state.
These terms vanish in the chirally symmetric phase when $m = 0$, or 
in the zero temperature limit when $\alpha_p\beta_p = \gamma_p\delta_p = 0$. 
However Eq.~(\ref{eq:HOFF1}) {\em can not vanish for finite $T$ in the broken
symmetry phase.}

In the standard treatment of BCS theory of superconductivity \cite{callaway}
the gap equation is obtained by demanding that the second order off--diagonal
Hamiltonian vanishes. We could also have followed this path and derived the 
gap equation from ${\cal H}_{{\rm OD}}$.
Instead we used the result obtained from the equation of motion to simplify
the second order off--diagonal Hamiltonian as a consistency check. Also,
in the BCS theory the condensation takes place between particles and holes in
the spin {\em singlet} state and thus contributions from spin {\em triplet}
terms in the Hamiltonian are usually ignored \cite{callaway}.
In the present case the chiral condensate is also a spin singlet bound state
and if we adapt an analogous approximation Eq.~(\ref{eq:HOFF1}) can
also be ignored.

\section{SU(2)$_{\rm f}$ $\otimes$ SU(3)$_{\rm c}$ NJL model at finite $T$
and $\mu$}
\label{sec5}
We now apply the method presented in the preceeding sections to the
SU(2)$_{\rm f}$ $\otimes$ SU(3)$_{\rm c}$ NJL model consisting of
scalar $(0^+, T = 0)$ and pseudoscalar $(0^+, T = 1)$ color singlet
$\bar{q}q$ channels as well as scalar and axial vector $(1^+, T = 1)$
color $\bar{3}$ $qq$ channels. This model was used in \cite{ish95} to study 
three quark bound states in the NJL model and is defined as
\begin{eqnarray}
{\cal L} 
& = &  
i\bar{\Psi} \partial\!\!\!/ \Psi 
+ g_1 \Bigl[ (\bar{\Psi} \Psi )^2 + (\bar{\Psi} i\gamma_5 \vec{\tau} \Psi)^2  \Bigr] 
+ g_2 \frac{3}{2}\Bigl[ (\bar{\Psi} \gamma_5 C \tau_2 \lambda^A \bar{\Psi}^T )
(\Psi^T C^{-1}\gamma_5 \tau_2 \lambda^A \Psi ) \Bigr]
\nonumber \\
&   & \;\;\;\;\;\;\;\; 
+\; g_3 \frac{3}{2} \Bigl[ (\bar{\Psi} \gamma_\mu C \tau_a\tau_2 \lambda^A \bar{\Psi}^T )
(\Psi^T C^{-1}\gamma^\mu \tau_2\tau_a \lambda^A \Psi ) \Bigr]
\label{eq:SU2SU3}
\end{eqnarray}
where $\lambda^A$ with $A = 2, 5, 7$ project out the color $\bar{3}$ channel.\footnote{The
Gell--Mann matrices $\lambda^a$ are normalized as
${\rm Tr}(\lambda^a\lambda^b) = 2\delta_{ab}$.} Summation convention is
implied for repeated indices. Positive values for the coupling
constants $g_1$, $g_2$ and $g_3$ implies
attraction in color $1$ and $\bar{3}$ channels. 

Just as in the case of the U(1) NJL model we Haag expand the Heisenberg field operators 
and apply the thermal Bogoliubov transformation to
the first order terms. Use of the equation of motion corresponding to 
Eq.~(\ref{eq:SU2SU3}) immediately yields the total chemical potential and
the gap equation for finite $T$ and $\mu$
\begin{eqnarray}
\mu
& = &
\mu_0 - 4 g_1 \int d^3\!\bar{p}\;\Bigl(\beta_p^2 - \delta_p^2 \Bigr)
\label{eq:COLORCHEM} \\
m 
& = & 
26 g_1 \int d^3\!\bar{p}\;\frac{m}{\omega_p} \Bigl(1 - \beta_p^2 - \delta_p^2 \Bigr)
\label{eq:COLORGAP} 
\end{eqnarray}
These results for $\mu$ and the gap equation are identical to the
Hartree--Fock results for the SU(2)$_{\rm f}$ $\otimes$ SU(3)$_{\rm c}$ NJL
model {\em without} the diquark channels \cite{reviews}. Indeed, we find that
to the lowest order in the Haag expansion the diquark channels do not
contribute to the equation of motion.

The diquark channels do contribute to the vacuum energy density which, with
the use of Eq.~(\ref{eq:COLORCHEM}) and (\ref{eq:COLORGAP}), can be
written as 
\begin{eqnarray}
\epsilon(T,\mu)
& = & 
-6 \Biggl\{12g_1\int d^3\!\bar{p}d^3\!\bar{q} + 2\int d^3\!\bar{p}\;\omega_p\; 
(1 - \beta_p^2 - \delta_p^2) 
- (26g_1 + 6g_2 - 72g_3) \left( \frac{m}{26g_1}\right)^2  \nonumber\\
&   & 
\;\;\;\;\;\;\;\;\;\;\;\;\;\;\;\;\;\;\;\;
+\;  \frac{(\mu - \mu_0)^2}{4g_1} 
- (6g_2 - 36g_3)
\left[\int d^3\!\bar{p}\;\Bigl(\beta_p^2 - \delta_p^2 \Bigr)\right]^2\Biggr\}  
\label{eq:COLORVAC}
\end{eqnarray}
Above expression reduces to the known vacuum energy density for the
SU(2)$_{\rm f}$ $\otimes$ SU(3)$_{\rm c}$ NJL model when
$g_2 = g_3 = 0$ \cite{reviews}. Similarly, the second order off--diagonal Hamiltonian 
can be simplified to the following form

\begin{eqnarray}
{\cal H}_{{\rm OD}} 
& = & m \sum_{l,\epsilon} \int d^3\!\bar{p}\; \Biggl\{
\sum_s s \Biggl[ 1 + \frac{1}{26 g_1}(6g_2 + 72g_3)\frac{m^2}{\omega_p^2} \Biggr]
\nonumber \\
&   & 
\otimes\Biggl[
\alpha_p\beta_p \tilde{B}^{\dagger}_{l\epsilon}(\vec{p},s) B^{\dagger}_{l\epsilon}(-\vec{p},s) 
+ \gamma_p\delta_p \tilde{D}^{\dagger}_{l\epsilon}(-\vec{p},s)D^{\dagger}_{l\epsilon}(\vec{p},s) 
\Biggr]
\nonumber \\
&   & 
\;\;\;\;\;\;\;\;\;\;\;\;\;\;\;\;\;
-\sum_s s \frac{1}{\omega_p}\Biggl[ \mu + (6g_2 + 36g_3) 
\int d^3\!\bar{p}\;\Bigl(\beta_p^2 - \delta_p^2 \Bigr)\Biggr]
\nonumber \\
&   & 
\otimes\Biggl[
\alpha_p\beta_p \tilde{B}^{\dagger}_{l\epsilon}(\vec{p},s) B^{\dagger}_{l\epsilon}(-\vec{p},s) 
- \gamma_p\delta_p \tilde{D}^{\dagger}_{l\epsilon}(-\vec{p},s)D^{\dagger}_{l\epsilon}(\vec{p},s) 
\Biggr] 
\nonumber \\
&   & 
\;\;\;\;\;\;\;\;\;\;\;\;\;\;\;\;\;
-\sum_{s_1, s_2} \Biggl[ \frac{1}{26 g_1}(6g_2 + 72g_3)\frac{m^2}{\omega_p^2} \Biggr]
\nonumber \\
&   & 
\otimes
\Biggl[
\alpha_p\gamma_p D^{\dagger}_{l\epsilon}(\vec{p},s_2)
B^{\dagger}_{l\epsilon}(-\vec{p},s_1) 
- s_1 s_2 \beta_p\delta_p \tilde{D}^{\dagger}_{l\epsilon}(-\vec{p},s_1)
\tilde{B}^{\dagger}_{l\epsilon}(\vec{p},s_2) 
\Biggr]
\nonumber \\
&   & 
\;\;\;\;\;\;\;\;\;\;\;\;\;\;\;\;\;\;\;\;\;\;\;\; \otimes 
\Bigl( \chi_1 \vec{\sigma}\cdot\vec{p} \chi_2 \Bigr) \Biggr\}
\label{eq:COLOROFF}
\end{eqnarray}
where $l = u, d$ and $\epsilon = 1, 2, 3$ are the flavor and color indices,
respecitvely and $\chi_i$ is the two component spinor for spin $s_i$.

One sees immediately that the Hamiltonian for the
SU(2)$_{\rm f}$ $\otimes$ SU(3)$_{\rm c}$ NJL model is exactly
diagonalized in the Breit--Wigner phase when $m = 0$. However
in the broken symmetry phase the first two terms in
Eq.~(\ref{eq:COLOROFF}), which is similar to Eq.~(\ref{eq:HOFF1}), vanish
only in the zero temperature limit but the third term in general does
not. This term consists of
particle--anti--particle and hole--anti--hole excitations with zero total three
momentum with the spins coupled to both 0 and 1. Since it
contains spin singlet excitations it is desirable that this term
vanishes exactly in order to obtain a true ground state of the
many--body system at least in the zero temperature limit. This can be accomplished 
if we impose the condition $g_3 = -\frac{1}{12} g_2$ on the coupling constants for
the scalar and axial vector diquark interactions. An immediate
consequence of this restriction is that the scalar and axial vector $qq$ channels
can not both be attractive or repulsive. 

In Figure~1a we plot the phase transition curve for the
SU(2)$_{\rm f}$ $\otimes$ SU(3)$_{\rm c}$ NJL model using the
Hartree approximation for the $\bar{q}q$ channel.\footnote{In this
approximation Eq.~(\ref{eq:COLORCHEM}) becomes $\mu = \mu_0$ and the 
coefficient infront of the integral in Eq.~(\ref{eq:COLORGAP}) changes
from 26 to 24.} The input
parameters for all our numerical work are $\Lambda = 0.05$
GeV, $g_1 = 5.01$ GeV$^{-2}$ and $g_2 = -12 g_3 = 3.11$ 
GeV$^{-2}$ which are taken from \cite{sch99}. On the phase transition 
line $m=0$ and the model Hamiltonian is exactly diagonalized. 
The corresponding vacuum energy densities, 
$\epsilon(T, \mu) - \epsilon(0, 0)$, are shown in Figure~1b. 
The upper curve corresponds to the energy density obtained with only 
the $\bar{q}q$ channel while the lower one includes both the $\bar{q}q$ and $qq$ 
channels. For all values of temperature and 
chemical potential on the phase transition line the inclusion of the 
scalar and axial vector diquark channel lowers the vacuum energy 
density of the system.

The effect of the diquark channel on the energy density is similar
in the broken symmetry phase at finite temperature. Figure~2a shows
the behaviour of the asymptotic mass $m$ as a function of $\mu$ for
$T = 0.1$ GeV. The phase transition is second order, which is a
consequence of the Hartree approximation for the $\bar{q}q$ channel,
and the critical chemical potential is found to be $\mu_{\rm c} = 0.253$ GeV.
Below $\mu_{\rm c}$ the Hamiltonian is only approximately diagonalized
and in Figure~2b we plot the vacuum energy density as a function of $\mu$.
As in Figure~1b the upper curve is obtained with only the $\bar{q}q$ channel
while the lower curve has contributions from both the $\bar{q}q$ and
$qq$ channels. Note that in both cases the values of the vacuum energy
densities are largest at $\mu = \mu_{\rm c}$ indicating that the broken
symmetry phase is the energetically favored phase. We also find that the
addition of the $qq$ channel lowers the vacuum energy density for finite 
density at zero temperature. Hence the inclusion of the diquark channel can 
lower the vacuum energy density of the system when the Hartree approximation 
is invoked for the $\bar{q}q$ channel.

\section{Summary and Outlook}
\label{concl}
In this work we have extended the NQA to quantum field theory to finite 
temperature and chemical potential. The basic idea is to subject the
asymptotic fields in the Haag expansion to a thermal Bogoliubov transformation
accompanied by a thermal doubling of the Hilbert space. Temperature and chemical 
potential are introduced non--linearly through the coefficients 
of the transformation. The NQA allows one to solve the operator equations of 
motion without explicitly specifying the structure of the ground state although it 
can be obtained a posteriori by diagonalizing the Hamiltonian. All quantities of
interest are determined through the algebra of asymptotic fields. One only needs to
assume the existence of a ground state which is annihilated by the
annihilation operators used in the Haag expansion. The thermal doubling of the 
Hilbert space is implicit in Eq.~(\ref{eq:AVAC}) where a thermal ground state was 
introduced which is annihilated by the thermal annihilation operators.

We tested our formalism on the U(1) and SU(2)$_{\rm f}$ $\otimes$
SU(3)$_{\rm c}$ versions of the NJL model and found that the first
term in the Hagg expansion can reproduce the mean field results for the number
and scalar densities, the chemical potential, the gap equation and the
vacuum energy density for finite $T$ and $\mu$.
Furthermore we found that the second order Hamiltonian can not be diagonalized 
at finite $T$ when the models are in the broken symmetry phase. 

When scalar and axial vector $qq$ channels are added to the SU(2)$_{\rm f}$ $\otimes$
SU(3)$_{\rm c}$ NJL model we found it necessary to impose a relation
on the scalar and axial vector coupling constants to diagonlize the
Hamiltonian for zero temperature. As a consequence scalar and vector
diquark channels can not simulateously be attractive or repulsive. 
For an attractive scalar diquark channel and in the Hartree approximation to
the $\bar{q}q$ channel, this restriction can lead to the lowering of the
vacuum energy density. These findings should be taken into account when
the work developed in \cite{ish95} to study three quark bound states in the
NJL model is extended to finite $T$ and $\mu$.

To go beyond the mean field approximation it is necessary to add higher order 
terms in the Haag expansion as shown in Eq.~(\ref{eq:HAAGEX}).
Here the second order terms involve both fermionic and bosonic fields,
the latter corresponding to various bound states in the model. It is
necessary to subject both types of fields to their respective thermal
Bogoliubov transformations accompanied by a redefinition of the
thermal ground state so that it is annihilated by both the
fermionic and bosonic annihilation operators. In the NJL model it is possible to
analytically calculate the bound state amplitudes by exploiting their
behaviour under parity, time--reversal and charge--conjugation transformations
\cite{gre86}. Once these amplitudes have been determined for finite $T$ and
$\mu$ they can be fed back into the Haag expansion of the Heisenberg fields 
and quantities of interest recalculated.

We emphasize that the formalism presented here is quite general and
has wide applications. Recently the importance of the scalar diquark
channel in the study of color superconductivity has been emphasized
by several authors using an effective four--Fermion instanton induced 
interaction\cite{diquark}. Our method is very well suited for
this application and investigation into the phenomenology of color superconductivity 
at finite temperature and chemical potential beyond the mean field
approximation is currently in progress. Other applications include studies
of pion or kaon condensations or properties of pion gas in chiral perturbation 
theory. In fact, our method can be applied to any quantum field theory in Minkowski 
space with well defined asymptotic fields. 
\acknowledgements
This work is supported in part by SEUIYD--PB97--1227. Y.~U. is also supported 
by the fellowship of Ministerio de Education y 
Ciencia de Espa\~{n}a under the auspices of the program "Estancias 
Temporales de Cientificos y Tecn\'{o}logos Extranjeros en Espa\~{n}a".
%
% BIBLIOGRAPHY

%
\vfill\eject
%
% FIGURES
%
\begin{center}
\begin{figure}
\epsfxsize=15cm
\epsfysize=16.5cm
\epsffile{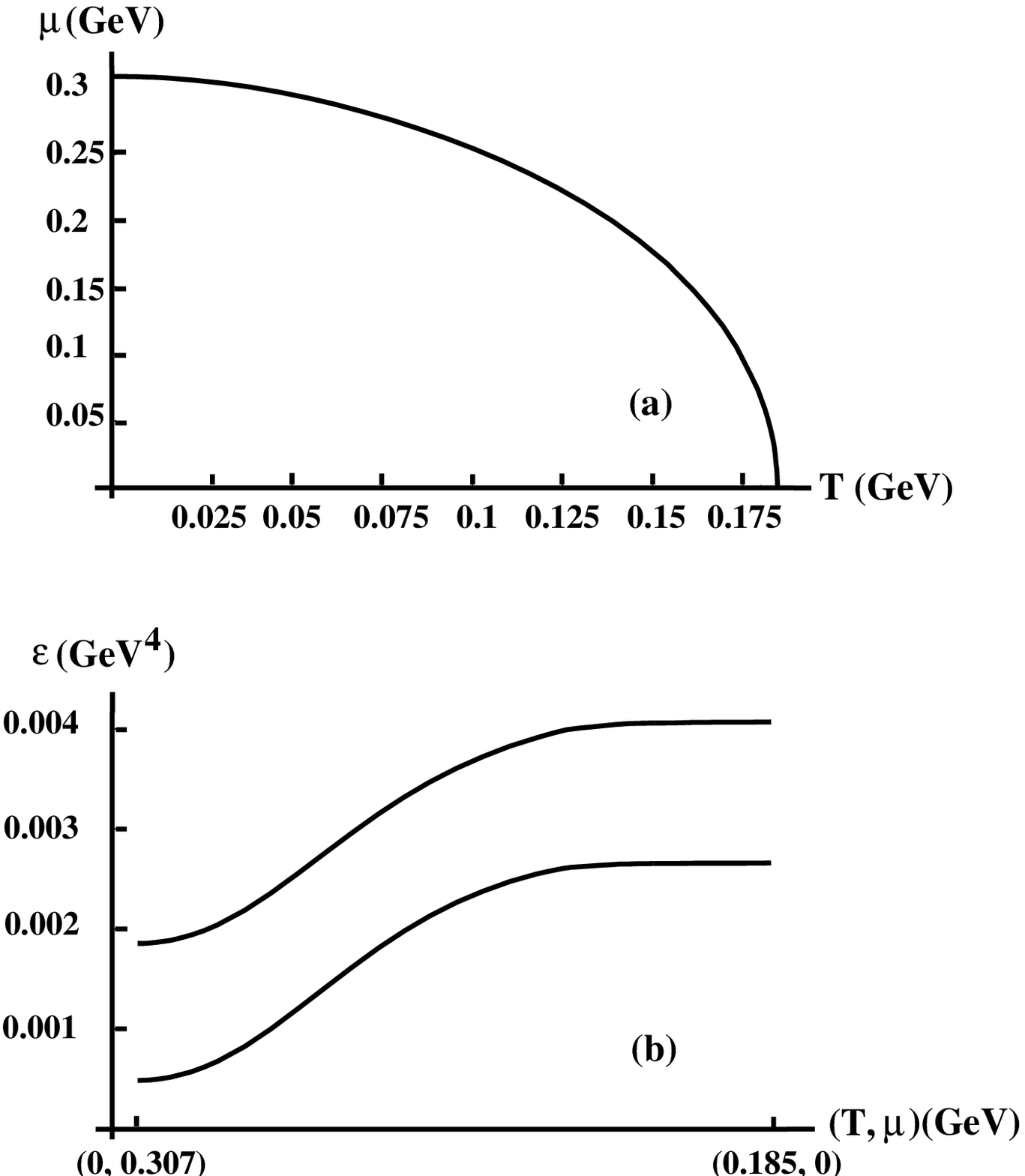}
\vskip 1cm
\caption{a) Phase trasnsition line for the
SU(2)$_{\rm f}$ $\otimes$ SU(3)$_{\rm c}$ NJL model with $\Lambda = 0.05$ GeV and
$g_1 = 5.01$ GeV$^{-2}$.
b) Vacuum energy density $\epsilon(T, \mu) - \epsilon(0, 0)$ on the phase
transition line begining at $(T, \mu) = (0, 0.307)$ GeV and
ending at $(T,\mu) = (0.185, 0)$ GeV. The upper and lower
curves have been obtained
with $g_1 = 5.01$ GeV$^{-2}$, $g_2 = g_3 = 0$ and $g_1 = 5.01$ GeV$^{-2}$, 
$g_2 = -12 g_3 = 3.11$ GeV$^{-2}$, respectively. }
\end{figure}
\end{center}
\begin{center}
\begin{figure}
\epsfxsize=15cm
\epsfysize=16.5cm
\epsffile{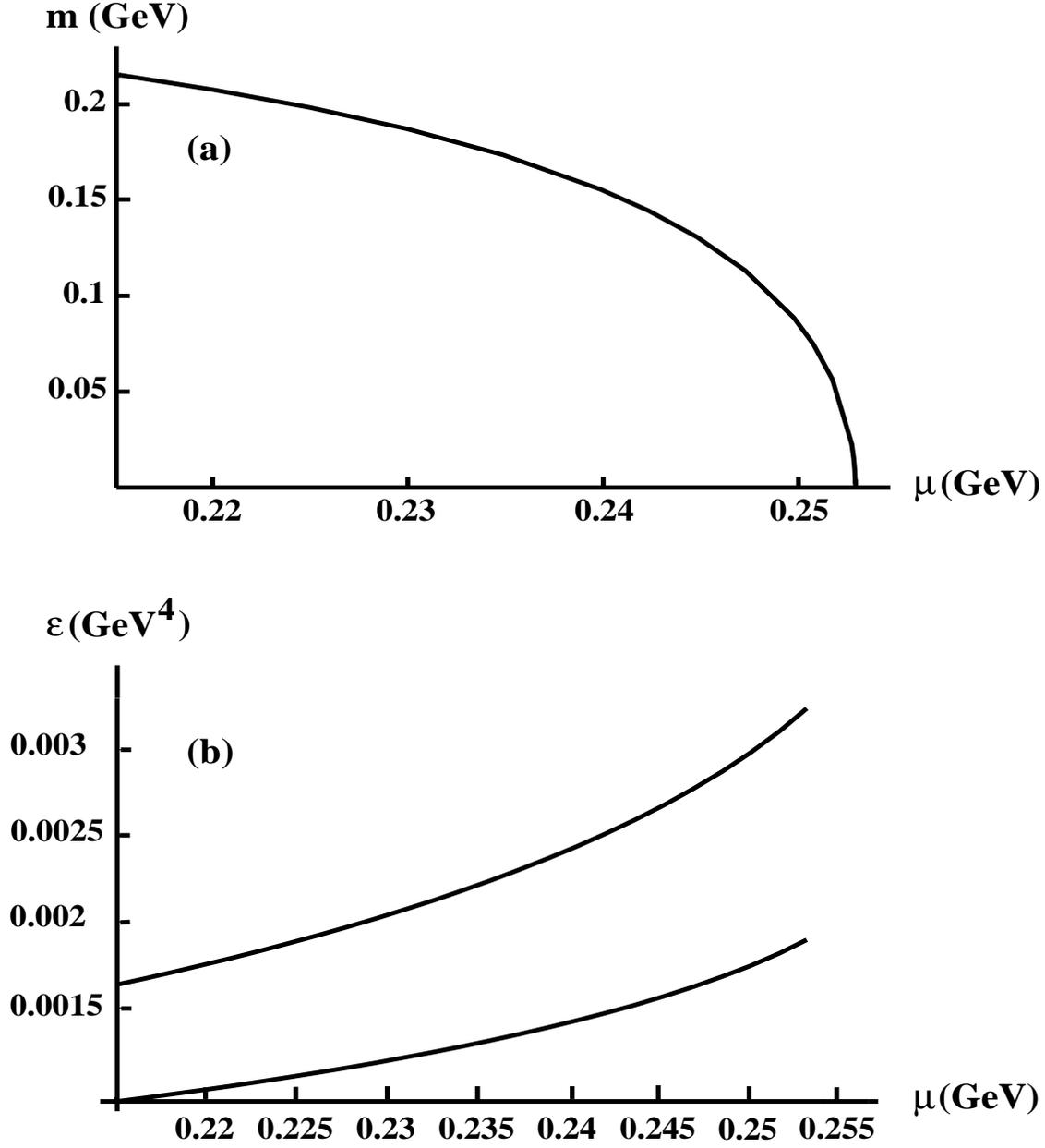}
\vskip 1cm
\caption{a) Asymptotic mass $m$ in the 
SU(2)$_{\rm f}$ $\otimes$ SU(3)$_{\rm c}$ NJL model as a function of chemical 
potential $\mu$ for $T = 0.1$ GeV with the same model parameters as in Figure~1a.
b) Vacuum energy density $\epsilon(T, \mu) - \epsilon(0, 0)$ as
a function of $\mu$ for $T = 0.1$ GeV. The upper and lower curves
have been obtained 
with $g_1 = 5.01$ GeV$^{-2}$, $g_2 = g_3 = 0$ and $g_1 = 5.01$ GeV$^{-2}$, 
$g_2 = -12 g_3 = 3.11$ GeV$^{-2}$, respectively.}
\end{figure}
\end{center}
\end{document}